\documentclass[twocolumn,showpacs,preprintnumbers,amsmath,amssymb]{revtex4}

\usepackage{amsmath,amssymb,color,epsfig,latexsym,natbib,graphicx,dcolumn}
\usepackage{bm}
\usepackage{ulem}
\bibpunct{[}{]}{,}{n}{,}{,}             
  
\newcommand{\bB}{\textbf{B}}

\newcommand{\bk}{\textbf{k}}

\newcommand{\rem}[1]{}

\newcommand\vecp[1]{\vec{#1}}                   

\newcommand{\be}{\begin{equation}}
\newcommand{\ee}{\end{equation}}

\def\bB0{\vecp{B}_0}   

\begin{document}

\title{\bf Where we observe that helical turbulence prevails over inertial waves \\
 in forced rotating flows at high Reynolds and low Rossby numbers}
\author{J. Baerenzung
{\footnote{presently at IMTG, Grenoble, France.}}}
\author{D. Rosenberg}
\author{P.D. Mininni
{\footnote{also at Departamento de F\'{\i}sica, Facultad de Ciencias Exactas y Naturales, Universidad de Buenos Aires and CONICET, , Ciudad Universitaria, 1428
         Buenos Aires, Argentina.}}}
\author{A. Pouquet}
\affiliation{NCAR, P.O. Box 3000, Boulder, Colorado 80307-3000, U.S.A.}

\begin{abstract}

We present a study of spectral laws for helical turbulence in the presence of solid body rotation up to Reynolds numbers $Re\sim 1 \times 10^5$ and down to Rossby numbers $Ro\sim 3 \times 10^{-3}$. The forcing function is a fully helical flow that can also be viewed as mimicking the effect of atmospheric convective motions. We test in the helical case variants of a model developed previously (Baerenzung et al. 2008a) against direct numerical simulations (DNS), using data from a run on a grid of  $1536^3$ points; we also contrast its efficiency against a spectral Large Eddy Simulation (LES) (Chollet and Lesieur 1981) as well as an under-resolved DNS. The model including the contribution of helicity to the spectral eddy dissipation and eddy noise behaves best, allowing to recover statistical features of the flow. An exploration of parameter space is then performed beyond what is feasible today using DNS. At fixed Reynolds number, lowering the Rossby number leads to a regime of wave-mediated inertial helicity cascade to small scales. However, at fixed Rossby number, increasing the Reynolds number leads the system to be dominated by turbulent energy exchanges where the role of inertial waves is to weaken the direct cascade of energy while strengthening the large scales. We find that a useful parameter for partitioning the data is $N_C=Re Ro=U^2_{rms}/[\nu \Omega]$, with $U_{rms}, \ \nu$ and $\Omega$ the rms velocity, the viscosity and the rotation rate respectively. The parameter that determines how much the energy cascade is direct or inverse--in which case the cascade to small scales is predominantly that of helicity--is linked to $Ro$.
\end{abstract}
\maketitle

\section{Introduction}

Turbulent flows at high Reynolds number prevail in the atmosphere and oceans, as well as in astrophysical settings, e.g. in solar-terrestrial (space weather) interactions. Because of the limited power of computers, even with the petascale efforts presently under way, modeling of such flows is needed in order to take into account the effect that the unresolved small scales, the so-called subgrid scales, have on the large scales, as done for example in the case of numerical weather prediction (Palmer 2001). Numerous models have been put forward and tested over the years, with, as their principal ingredients, an eddy viscosity representing the dissipation of energy linked to the unresolved small-scale eddies (Smagorinsky 1963, Kraichnan 1976), and an eddy noise (Leith 1971) which mimics the stochasticity of the small scales (see Meneveau and Katz 2000 for a recent review).

In the presence of helicity (velocity-vorticity correlations) nonlinear terms are weakened. This was invoked by Lilly (1986) (see also Anthes 1982) to explain the persistence of strong storms, as observations of helical features are common in atmospheric flows, e.g. when analyzing VORTEX data (Verification of the Origin of Rotation in Tornadoes EXperiment, Markowski et al. 1998). Tornadoes can be encountered in two types of structures: one that is laminar (like water spouts) and one that is strongly turbulent, typical of storms in the central USA. To study the latter, two-dimensional incompressible studies of axi-symmetric tornado-like vortices in the presence of vertical forcing were performed in cylindrical coordinates with resolutions of $64^2$ points using eddy viscosity (Nolan and Farrell 1999).; an analysis of 120 simulations in terms of the Reynolds ($Re$) and Rossby ($Ro$) numbers indicated that the determinant parameter was $Re/Ro=\Omega L^2/\nu_{\ast}$ with $\Omega$ the imposed rotation, $\nu_{\ast}$ the eddy viscosity and $L$ a characteristic scale. In such a study, the rotation is imposed, whereas in a tornado the rotation has a somewhat uncertain origin (Rotunno 1984, Rotunno and Klemp 1985): it is thought to be linked to the pre-existence of a downdraft which, together with precipitation introduced as in Markowski et al. (2003), transports angular momentum to the ground where the circulation is then closed (see Wicker and Wilhelmson 1995, and references therein). More recent studies of tornado genesis have included the effects of moisture and buoyancy, together with stratification and latent-heat release; this leads to a more realistic modeling of supercell storms and of their maximal vorticity.

A variety of physical phenomena interact to create the dynamics of such 
flow structures, among which 
are strong local rotation (induced for example by rain in downdrafts), strong winds with accelerations several times 
that of gravity, stratification, and boundary layer effects. Many experimental, observational, phenomenological, theoretical, 
and numerical studies have been devoted to these interactions and, as the power of 
instrumentation increases, 
progress will continue to be made. Crucially, such flows and flow structures are embedded in an atmospheric circulation 
at very high Reynolds number,
leading to the formation of a myriad of small-scale intense structures and to a turbulent, unpredictable background 
flow. As usual, two approaches to studying these flows can be contrasted: on the one hand, models as complete 
as possible must be devised and analyzed in a parametric fashion. On the other hand, a reductionist approach, 
as in applied mathematics, calls for drastic simplifications,  as in the case, for example, of the studies that 
considered the tornado as an axi-symmetric two-dimensional feature which were mentioned earlier. 
 
Here, we propose to reduce the problem in a different way by considering the related issue of hydrodynamic rotating 
turbulence, ignoring the thermodynamics, but considering highly resolved three-dimensional features that can help 
determine turbulent properties that may be relevant for complex atmospheric flows, and insisting on the 
importance of attaining and modeling as high a Reynolds number as possible with, at the same time, realistic 
Rossby numbers. In Mininni and Pouquet (2009bc), using a direct numerical simulation (DNS hereafter) on 
$1536^3$ points of rotating turbulence forced with a helical flow, it was shown that the spectral indices 
for the energy and helicity spectra differ from the classical Kolmogorov law. The purpose of this paper is 
to reach higher Reynolds numbers than what was attained in this massive DNS, as well as lower Rossby numbers. 
Our motivation is to understand the effect of rotation on the flow, the impact of Rossby waves on scaling laws,
and to quantify the effect of helicity. To that effect, we shall first assess the validity of several models 
against this DNS, based on previous work for non-rotating turbulence (Baerenzung et al. 2008a; see also Baerenzung 
et al. 2008b for the rotating non helical case, and Baerenzung et al. 2009 for the non-helical case of coupling to 
a magnetic field). Armed with this model, we shall examine the variation of spectral indices for the energy and 
helicity cascade to small scales as a function of both the Reynolds and Rossby numbers, and using a parametric
study, we shall identify different behaviors depending on the dimensionless parameters of the problem. The next 
section gives the basic equations, and \S \ref{s:testing} analyzes data on temporal evolution and spectra stemming 
from the models and the DNS. Then, \S \ref{s:param} presents the parametric study, and finally 
\S \ref{s:conclu} presents our conclusions.

\section{Equations and models}\label{s:eqs}

The incompressible ($\nabla \cdot \bm v = 0$) Navier-Stokes equations for a flow corresponding to dry dynamics, with velocity $\bm v$ and constant density $\rho_0\equiv 1$ read, 
\begin{equation}
\frac{\partial {\bf v}}{\partial t} + {\bf v} \cdot \nabla {\bf v} + 2 \Omega \times {\bf v} 
= -\nabla P + \nu \Delta {\bf v} +{\bf F^v} \ ,
\label{eq:NS}
\end{equation}
where $P$ is the total pressure (including the centrifugal force), $\mbox{\boldmath $\Omega$}$ is the imposed solid body rotation, taken to be in the $z$ direction, and ${\bf F^v}$ is a forcing term mimicking the input of energy (and helicity) to the system through, e.g., buoyancy forces. In the absence of viscosity and forcing ($\nu\equiv 0,\ {\bf F^v}\equiv 0$), the energy $E=\left<{\bf v}^2/2\right>$ and the helicity $H=\left<{\bf v} \cdot \mbox{\boldmath $\omega$}/2\right>$ (Moffatt 1969) are conserved (with $\mbox{\boldmath $\omega$} = \nabla \times {\bf v}$ the flow vorticity). Conservation laws are thought to play an important role in the dynamics of turbulent flows. The role of helicity in the absence of rotation has been studied in a number of papers (see e.g. Brissaud et al. 1973, Kraichnan 1973, Andr\'e and Lesieur 1977, Hunt and Hussein 1991, Moffatt and Tsinober 1992, Holm and Kerr 2002, Kurien et al. 2004, Chen et al. 2005, Jacobitz et al. 2008, Krstulovic et al. 2009): the energy and helicity spectra ${E}(k,t)$ and ${H}(k,t)$ (averaged over spherical shells of width $\Delta k=1$), follow a Kolmogorov law with ${E}(k,t) \sim k^{-5/3}$ and ${H}(k,t) \sim k^{-5/3}$. 

Using the rms velocity, $U_{rms}$, the Reynolds and Rossby numbers are defined, respectively, from Eq. (\ref{eq:NS}) as
\begin{equation}
Re= U_{rms}L_0/\nu \ , \ \ \ \ \ \ \  Ro= U_{rms}/2L_0\Omega \ ,
\label{REY} \end{equation}
where $L_0 = 2\pi \int{E(k) k^{-1} dk}/\int{E(k) dk}$ is the isotropic integral length scale. We can also evaluate a micro Rossby number
$Ro^{\omega}=\omega_{rms}/2\Omega $
where $\omega_{rms}$ is the rms vorticity, and which measures the strength of local compared to imposed rotation. 
 In all simulations, this number is close to unity or larger; here and in \ref{s:testing}, $\nu=1.6 \times 10^{-4}$. Note that in the presence of strong rotation, an inverse cascade of energy to large scales is observed and the total energy is found to grow with time. In this case, $Re$ and $Ro$ given in Table \ref{tab1} below are assigned with the turbulence statistics evaluated at the onset of the inverse cascade, $T_I$.

The computations we analyze are set in a cubic box of length $2\pi$ with minimum and maximum wavenumbers $k_{min}=1$ and $k_{max}=N/3$, $N$ being the number of grid points in each direction and using the usual 2/3 dealiasing rule. The code is pseudo-spectral with temporal integration using a second-order Runge-Kutta scheme.  The forcing, of amplitude $F_0$ 
 is given by a Beltrami flow:
\begin{eqnarray}
{\bf F^v} / F_0 &=& \left[B \cos(k_F y) + C \sin(k_F z) \right] \hat{x} + 
    \nonumber \\
&& {} \left[C \cos(k_F z) + A \sin(k_F x) \right] \hat{y} + \nonumber \\ 
&& {} \left[A \cos(k_F x) + B \sin(k_F y) \right] \hat{z} .
\end{eqnarray}
This so-called ABC flow, an eigenfunction of the curl with eigenvalue $k_F$ injects both energy and helicity in the system. Such flows give a periodic checkerboard of overturning rolls with jets at the circulation centers and were used in Lilly 1986 to mimic flows in rotating thunderstorms. In all runs we take $A=0.9$, $B=1$, $C=1.1$, and $k_F=7$, so as to cover both the direct cascade of energy and helicity to small scales and the inverse cascade of energy  to large scales.

In what follows, we consider a DNS run that has been performed on a grid of $N^3=1536^3$ points, the largest ever for rotating turbulence to our knowledge, and several Large Eddy Simulation (LES) runs. The LES tests presented in the next section were performed on a grid of $96^3$ points, while the parametric study of \S \ref{s:param} was done with LES up to $512^3$ grid points for the highest Reynolds number studied. A statistically steady state of turbulence with almost no rotation ($\Omega=0.06$) was first reached in the DNS, after approximately ten turn-over times $\tau_{NL}=L_0/ U_{rms}$. Then, rotation was switched on, with $\Omega$ between $1.75$ and $117$ for the parametric study, at a time taken as the initial condition (defined as $t=0$) for all the runs analyzed in this paper (DNS and LES). For the LES initialization on grids of $N_L^3$ points ($N_L<N$), the $1536^3$ data at $t=0$ was truncated at $k_{max}=N_L/2$ (note no de-aliasing is needed in an LES) and the computations lasted roughly another thirty turn-over times.

Since the runs use a pseudo-spectral scheme with a spectral LES, we write the Navier-Stokes equation in terms of the velocity in Fourier space:
 $$(\partial_t + \nu k^2 )v_\alpha(\textbf{k},t)+ 2\Omega P_{\alpha\beta} \ \varepsilon_{\beta z \gamma}v_\gamma(\bk,t)= 
 t_\alpha^v(\textbf{k},t) + F_\alpha^v(\textbf{k},t) , $$ 
with $\varepsilon_{\beta z \gamma}$ the usual antisymmetric tensor. $\textbf{t}^v_{\alpha}(\textbf{k},t)$ is a bilinear operator for energy transfer written as
\begin{equation}
t_\alpha^v(\textbf{k},t)=-iP_{\alpha \beta}(\textbf{k})k_ \gamma\sum_{\textbf{p}+ \textbf{q} = 
\textbf{k}}v_{\beta}(\textbf{p},t)v_{\gamma}(\textbf{q},t),
\end{equation}
and the projector $P_{\alpha \beta}(\textbf{k})=\delta_{\alpha\beta}-k_\alpha k_\beta/k^2$ enforces incompressibility ($k_\alpha v_\alpha = 0$); Einstein summation 
is used. From the equation above, one builds the temporal evolution for the energy and helicity spectra $E(k)$ and $H(k)$, averaged over spherical shells of radius $k=|{\bf k}|$:
\begin{eqnarray}
(\partial_t + 2\nu k^2 )E(k,t)& = &
T_{E}(k,t) +  F_{E}(k,t)  \ , \label{energy}\\
(\partial_t + 2\nu k^2 )H(k,t)& = &
T_{H}(k,t) +  F_{H}(k,t) \ , \label{helicity}
\end{eqnarray}
with $T_{E,H}(k,t)$ the energy and helicity nonlinear transfers at wavenumber $k$ and 
$ F_{E,H}(k,t)$ the forcing expressed in terms of its symmetric (energy) and non-mirror symmetric (helicity) parts. The transfer terms involve triple correlations between three Fourier modes 
$\textbf{v}(\textbf{k},t)$, $\textbf{v}(\textbf{p},t)$ and $\textbf{v}(\textbf{q},t)$ 
with
$\textbf{p} + \textbf{q} = \textbf{k}$, expressing the fact that the Fourier transform of the nonlinear term is a convolution. Hence, one speaks of triadic interactions between the modes, and detailed energy and helicity conservation occurs for each such triad. This is where the classical closure problem of turbulence arises: one now needs an equation for triple moments which will involve quadratic correlations, and so on. Many closure schemes have been devised over the years, one of the most successful in spectral space being the Eddy Damped Quasi-Normal Markovian model (EDQNM hereafter, see Orszag 1977). Within the framework of the closure, the exact transfer terms $T_{E,H}$ are replaced by approximations which we denote $\widehat T_{E,H}(k,t)$. The EDQNM closure was written for helical turbulence in Andr\'e and Lesieur (1977); it leads to a set of integro-differential equations for the energy and helicity spectra. 
The EDQNM approximation has been used in atmospheric studies, in applications to shear, rotating or stratified flows including non-Markovianized versions based on extensions of the Direct Interaction Approximation (see O'Kane and Frederiksen 2008 and references therein).

The EDQNM balance equations for both the energy and the helicity can be expressed in terms of emission and absorption. For the energy, the former is related to eddy noise (an inhomogeneous term involving energy at all pairs of modes ${\bf p}, {\bf q}$ such that ${\bf k}={\bf p} + {\bf q}$ as stated before); the latter, linear in $E(k)$, gives rise to eddy viscosity, which corresponds to a drain on the energy at mode $k$, although in some cases it can be negative.

The EDQNM closure in the helical case leads to two extra contributions to the transport coefficients as helicity contributes to eddy viscosity and eddy noise. Since, dimensionally, the eddy viscosity, $[\tilde \nu_t]$, in the energy equation 
can be written  $[\tilde \nu_t]=[\nu/k]$,  we can write
\begin{equation}
\partial_t E^{}(k)\sim -2\left( \nu + \nu_t \right) k^2E^{}(k)  -2 \tilde \nu_t k^2 H^{}(k) .
\label{nu_turb}\end{equation}
 This uses a short-hand and simplified notation to bring out the structure of the model (see Baerenzung et al. 2008a for details), which omits multiplicative coefficients as well as both the resolved scale contributions and the eddy-noise contributions. While the EDQNM eddy viscosity $\nu_t (k,t) \sim \int f_1(k,p,q) E(q)dpdq$ depends (through the function $f_1$) on an integral of the energy spectrum in the small scales and represents the drain of energy from the resolved scales due to the unresolved sub-grid scales, $ \tilde \nu_t (k,t) \sim \int f_2(k,p,q)H(q)dpdq$ gives the contribution of small-scale helicity through a different function $f_2$ The total eddy noise represents the effects of the small-scale helicity as well as the effects of small-scale energy on the large scales of the flow. 

In a previous paper (Baerenzung et al. 2009), we tested for rotating flows the non-helical version of the model, with all pseudo-scalar terms (denoted with a tilde) equal to zero. 
However, it was shown recently that when both sizable rotation and helicity are present, the small-scale cascade is dominated by helicity (Mininni and Pouquet 2009ab), whereas the energy mostly undergoes an inverse cascade, as is well known. The question then arises as to whether a helical model is needed for flows where helicity plays a central role in the direct cascade dynamics. To that effect, we contrast in the next section results stemming from three models on $96^3$ grids: the full helical model, hereafter labeled LES-PH, the non-helical model, hereafter LES-P, and the Chollet-Lesieur model, in short, CL or LES-CL.
For completeness, we also compare the $1536^3$ DNS data to an under-resolved DNS on a grid of $160^3$ points (which, using the 2/3 dealiasing rule, has $k_{max}=53$, slightly larger than for the LES runs).

The Chollet-Lesieur (1981) model introduces an eddy viscosity of the form, with $k_{cut}=N/2-1$ a cut-off:
\begin{equation}
\nu_{CL}(k,t)=C\nu^+(k,t)\sqrt{E(k_{cut},t)/k_{cut}} \ .
\end{equation}  
 The quantity $\nu^+(k,t)= 1+3.58(k/k_{cut})^8$ is a dimensionless cusp function; $\nu_{CL}(k,t)$ replaces $\nu_t$ in Eq. (\ref{nu_turb}) (with $\tilde \nu_t = 0$). The CL model was derived from the EDQNM equations using the fact that $C\sqrt{E(k_{cut},t)/k_{cut}}$ is the asymptotic expression of nonlocal transfer from subgrid to resolved scales and assuming a Kolmogorov spectrum extending to infinity. $C$ in our runs was adjusted with the Kolmogorov constant computed from the ABC flow resolved by a DNS run using $512^3$ grid points, {\it viz.} $C=0.14$.

\section{Testing of the helical model} \label{s:testing}

\begin{figure}[h!] 
\includegraphics[width=8.5cm, height=45mm]{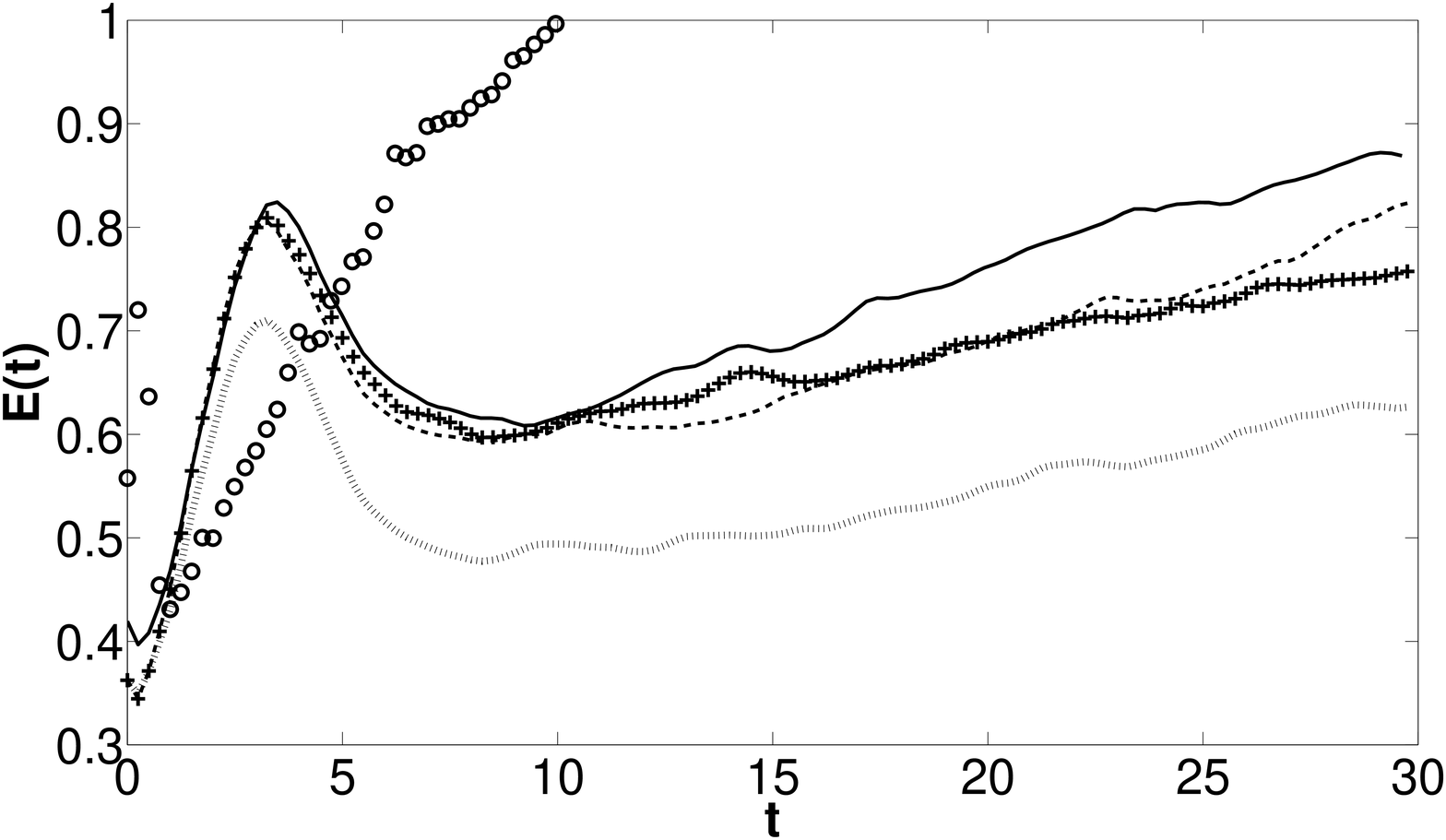}
\includegraphics[width=8.5cm, height=45mm]{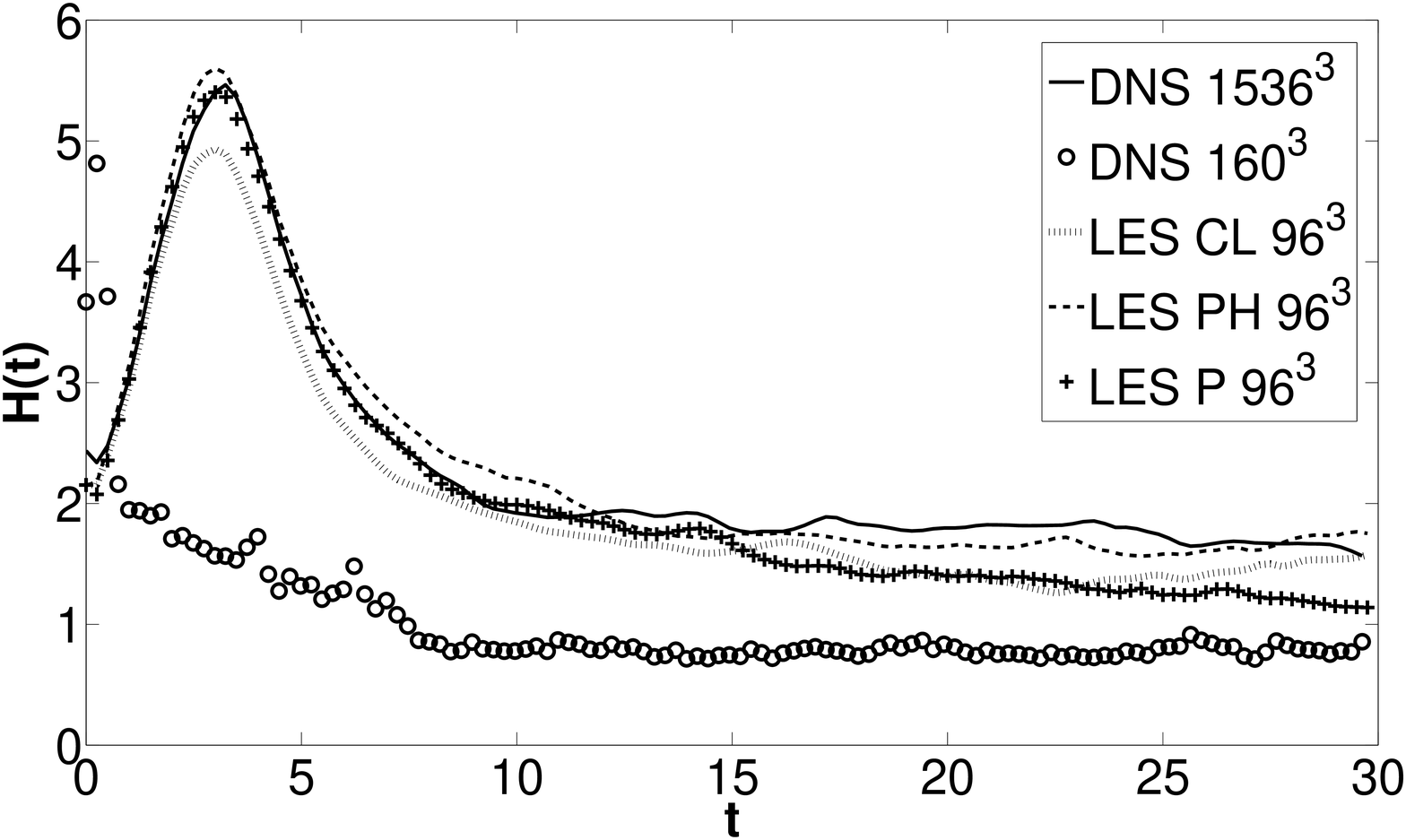}
\includegraphics[width=8.5cm, height=45mm]{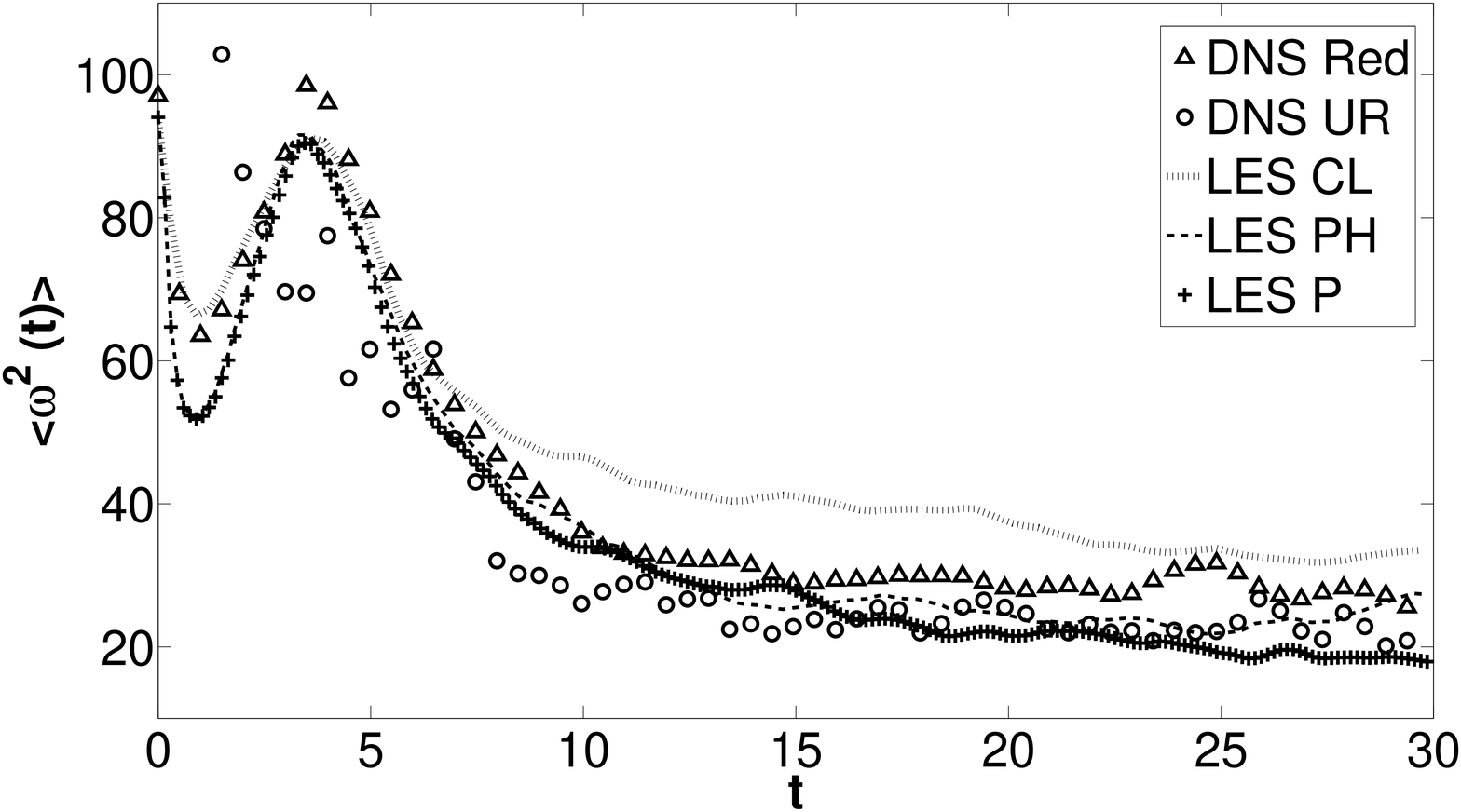}
\caption{
Evolution of the total energy (top), helicity (middle) and enstrophy (bottom) for several runs: DNS, $1536^3$ grid points (solid line), under-resolved DNS, $160^3$ points (circles), 
LES-CL (dotted line), LES-PH (dashes) and LES-P (crosses), all on $96^3$ points. The ``DNS Red'' (for ``reduced'') corresponds to the data of the $1536^3$ DNS filtered down to $96^3$ grid points. As a reference, the energy of the under-resolved simulation at $t=20$ is $E=1.39$, and at $t=30$ it is $1.69$.
} \label{compa_energy}  
\end{figure}

We show in Fig. \ref{compa_energy} a comparison of the temporal evolution of the total energy (top), helicity (middle) and enstrophy $\left< \omega^2 \right>$ (bottom) for five runs which were done specifically for comparison and validation purposes of the LES models against the high resolution DNS. 
They correspond to three $96^3$ LES runs (PH, P and CL) and an under-resolved DNS with the same $Re$ and $Ro$ as the $1536^3$ DNS; the reduced label stands for $1536^3$ DNS data filtered to $96^3$ points. In all that follows, the DNS is shown with a solid line, the LES with helical coefficients with a thick dotted line and that without the helical contributions with a thinner dash-dotted line, the Chollet-Lesieur model is shown with thin (grey) bars and finally the under-resolved run is displayed with circles. 
The energy displays three distinct temporal phases, all well reproduced by the models except for the under-resolved DNS), although to various degrees of accuracy. At first, the energy grows and then decays (up to $t=10$), due to an initial adjustment as the Coriolis force gets suddenly larger at $t=0$.
Even though the Coriolis force itself does not input energy in the system, inertial wave resonances render the flow quasi bi-dimensional and, after $t=10$, the subsequent growth of the energy is then attributed to the onset and further development of an inverse cascade of energy with a transfer to scales larger than the forcing scale $L_F$. The transition times between these phases are well reproduced by all models, and so is the growth rate in the inverse cascade, except for the LES-P model. The under-resolved run stands on its own: the lack of small-scale dissipation produces an unphysical growth of energy and its results, clearly unreliable, will not be commented upon further other than to say that with insufficient numerical resolution, a model indeed is needed to mimic the effects of the unresolved scales.

The helical model LES-PH is the one closest to the DNS, whereas the CL model is too dissipative, because it does not take into account the weaker nonlinearities due in part to the partial Beltramization of the flow. 
The temporal evolution of the total helicity follows the same pattern, except that all the injected helicity undergoes a direct cascade as can be verified by computing its flux (Mininni and Pouquet 2009b), and therefore its total value as a function of time attains a statistically steady equilibrium; when examining the errors (not shown) the LES-PH model is best overall, and the CL model is worst.
We also note that, at early time and up to the onset of the inverse cascade of energy for $t\sim 10$, the non-helical LES-P model behaves consistently better. This may indicate that as long as the inverse energy cascade has not begun and we are in a regime dominated by waves, the direct cascade to small scale is as much energy as it is helicity and thus the helicity in the model is not as essential. Overall, the normalized error for the total helicity using the LES-PH model is between  $10^{-3}$ and $10^{-1}$, reasonable values considering the very large difference in the linear grid resolution (by a factor of 16).

The comparisons between models and DNS for the total dissipation ${\cal D}(t)=2\nu\left<\omega^2(t)\right>$ (see Fig. \ref{compa_energy} bottom, noting all runs have the same $\nu$) show a large discrepancy which can be partially removed when filtering the DNS data down to the resolution of the LES runs. Dissipation
has its largest contribution from small scales which are not resolved in the LES runs
(see also Fig. \ref{scatter_tot} below for a comparison of dissipation between LES and unfiltered DNS data for different values of $\nu$ and $\Omega$ than those used here). When comparing to the filtered DNS displayed with triangles in Fig. \ref{compa_energy}, again LES-P and LES-PH give the best results, while the CL model over-estimates the dissipation by a consistent amount ($\approx 30\%$). We note that, in the case of the Smagorinsky model, a study of helical flows (Li et al. 2006) shows an under-estimation of helicity dissipation by 40\%. 

\begin{figure}[h!] 
\includegraphics[width=8.5cm, height=45mm]{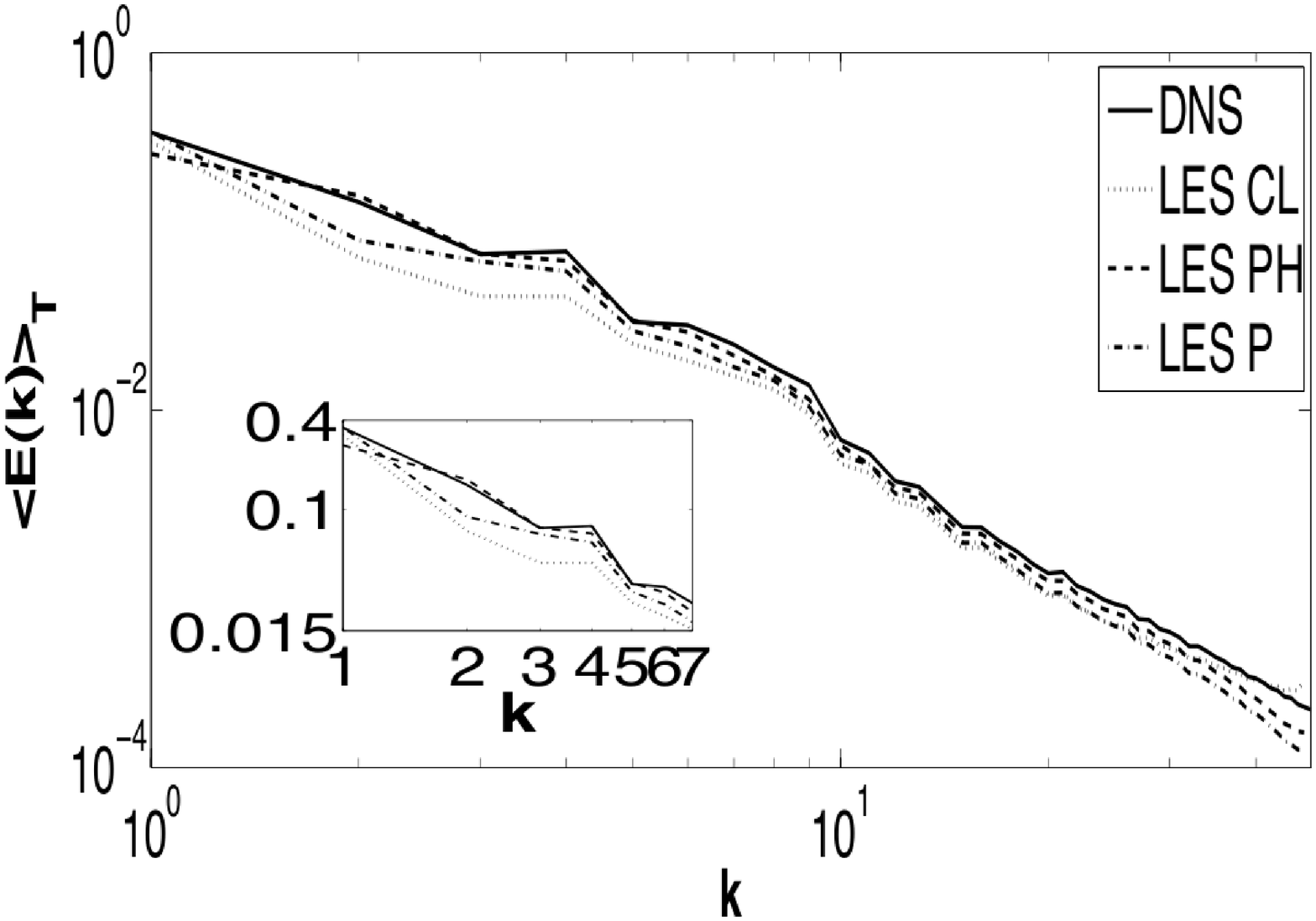}
\includegraphics[width=8.5cm, height=45mm]{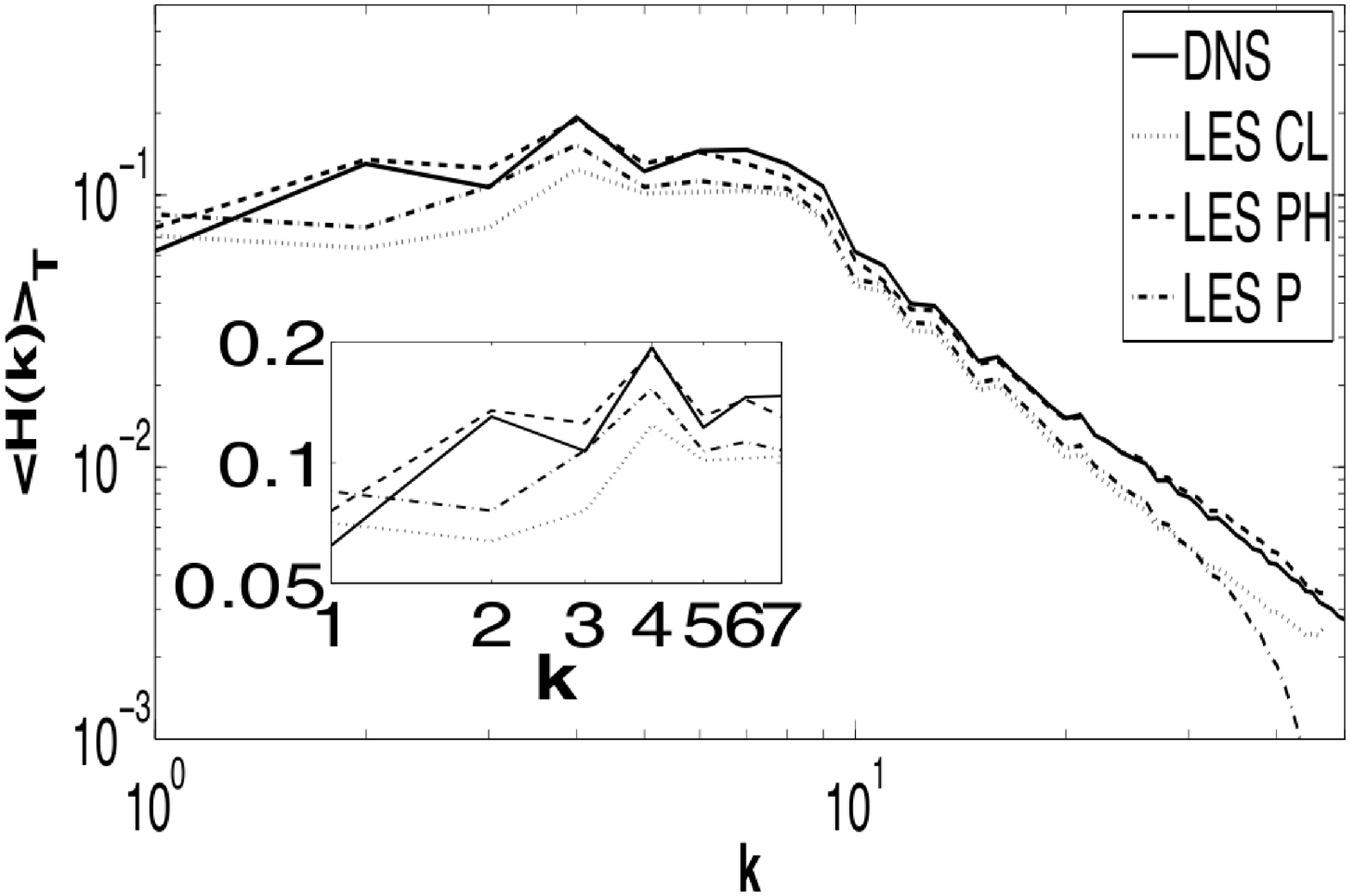}
\caption{Energy (top) and helicity (bottom) spectra averaged during the inverse cascade, from $t=20$ to $t=30$; same notation as in Fig. \ref{compa_energy}. Insets: zoom on the large scales; note the inadequacy of all models in this range of wavenumbers except for LES-PH, i.e. when taking into account the helical contributions to sub-grid modeling. 
} \label{compa_spec_T}  \end{figure}

When comparing spectra for energy and helicity, the same conclusions arise, with a better behavior for the helical model LES-PH. We show in Fig. \ref{compa_spec_T} the spectra  for energy (top) and helicity (bottom), averaged over an interval of time after the start of the inverse cascade of energy. All spectra stop at the maximum wavenumber of the LES, $k_{max}^{LES}=48$ and do not show a dissipation range by construction (note, however, that in our approach, we keep the (bare) viscosity in the equations, that of the DNS run on the grid of $1536^3$ points). In the small scales, the LES-PH model reproduces very closely the DNS up to $k_{max}^{LES}$, particularly so for the helicity. In the inverse cascade of energy, a good agreement is obtained as well, with the CL model being less efficient, a feature already noted on the temporal evolution of the total energy. The helicity does not undergo an inverse cascade for any of the runs performed in this paper, but the agreement with the DNS spectra is less striking than for energy.

For scales larger than the forcing scale $L_F$, $E(k)$ appears to follow a classical Kolmogorov law. There is evidence in other work for much steeper power laws in the large scales of rotating turbulence, {\it viz.} $E(k)\sim k^{-3}$ (Smith and Lee 2005, see also Tran and Bowman 2003), when computing in periodic boxes with variable aspect ratio and using hyperviscosity. The differences may be linked to the effect that wave interactions have on large scales on the one hand, and to a subsequent direct cascade of enstrophy because of the bi-dimensionalization of the flow on the other hand (Smith et al. 1999). These results, which may also be Rossby dependent (Chen et al. 2005) are, however, beyond the scope of the present study which focuses on small-scale properties. The spectrum $H(k)$ at large scales is flat and no clear power law can be identified. Note that the realizability condition $|H(k)|\le kE(k)$ does not present a dynamical constraint for the helicity as $k\rightarrow 0$ in the inverse cascade.
 
Finally, we note that LES-PH reproduces well the type of structures observed in the flow, and in particular the spatial juxtaposition of laminar updrafts that are fully helical (Beltrami core vortices) and a tangle of small-scale vortex filaments ordered in columnar structures (Mininni and Pouquet 2009c).
The size and shape of eddies in the horizontal plane are correctly captured, even though phase information is partially lost because of the intrinsic stochasticity and lack of predictability of the flow.

\begin{figure}[h!] 
\includegraphics[width=8.7cm, height=107mm]{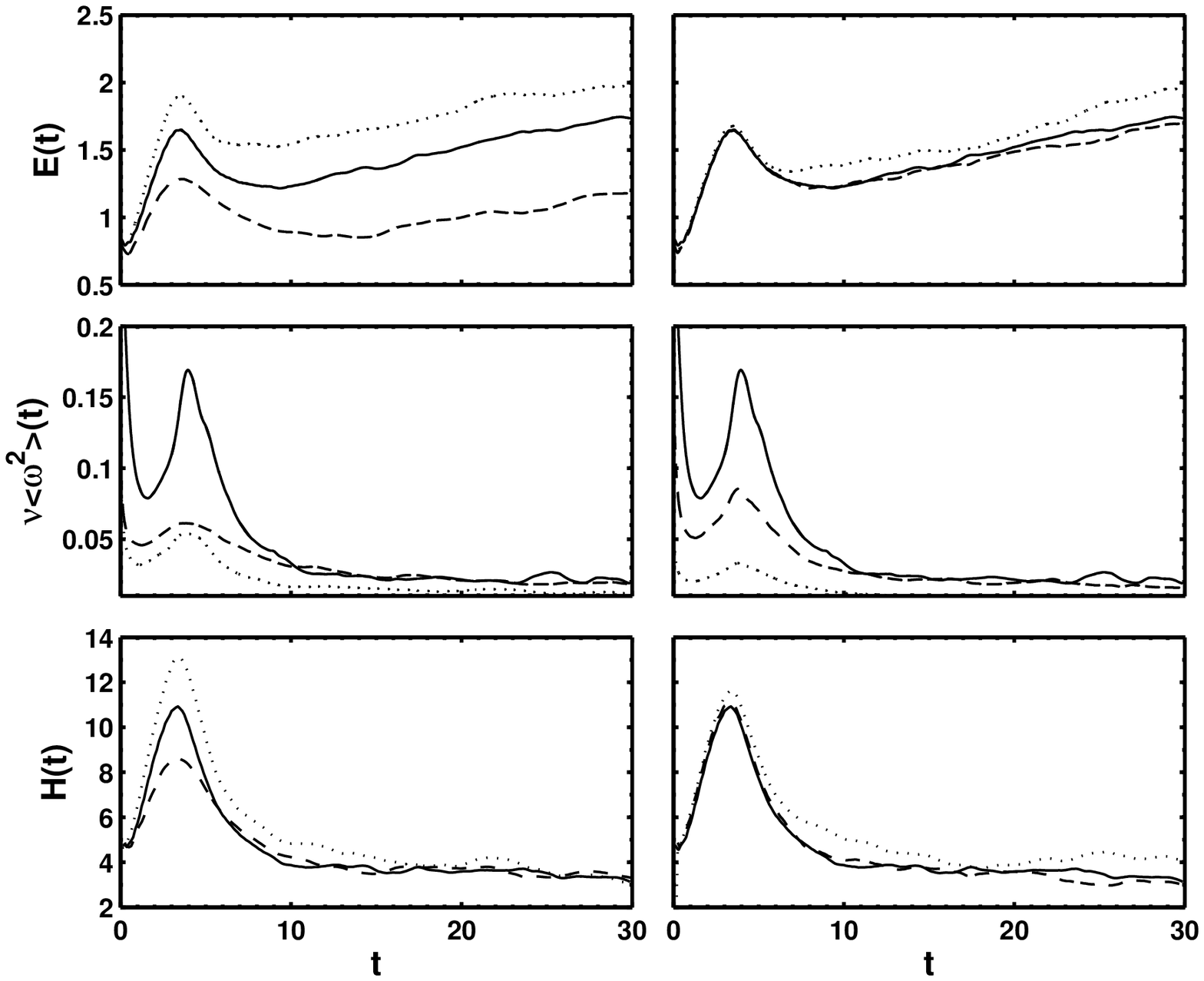}
\caption{Temporal evolution of energy (top), dissipation (middle) and helicity (bottom); the solid line in each plot represents the DNS run R5F ($\nu=1.6\times 10^{-4}$, $\Omega=9$), while all other runs are performed with LES-PH. {\it Left column:} Fixed viscosity ($\nu=1.6\times 10^{-4}$) with $\Omega$ varying between 4.5 and 18 by factors of 2 for runs R4 (dash line), R5F, and R6 (dotted line). {\it Right column:} Fixed rotation rate ($\Omega=9$) and $\nu$ decreasing from $2.5\times 10^{-4}$ to $8\times 10^{-5}$ for runs R2 (dash line), R5F, and R11 (dotted line).}
\label{scatter_tot}  
\end{figure}

\section{Turbulence versus waves} \label{s:param}

\begin{table*}
\caption{Parameters of the runs using the LES-PH model; $\nu$ is the viscosity, $\Omega$ the imposed rotation, $\omega_{rms}$ the rms vorticity at $T_I$, where $T_I$ is the time of onset of the inverse energy cascade; $Ro$ and $Re$ are the Rossby and Reynolds numbers, $U_{rms}$ the rms velocity, and $L_{O,IF}$ the integral scales at $T_I$ and $t=30$. Data is averaged $\in [T_I,T_I+5]$. Compared to their values at $T_I$, $U_{rms}$ has increased by roughly 20\% at t=29 and $\omega_{rms}$ has decreased by between 10\% and 20\%,  except for run R3 for which both are stationary, and for run R5F, for which $\omega_{rms}$ has decreased by 30\%. In the last three columns, the spectral index for the energy $e$, the sum of energy and helicity indices $e+h$, and the normalized flux ratio $\Pi_H/[k_F\Pi_E]$ are given at $T_I$. R3 and R5z have no discernible inverse cascade and should have a Kolmogorov spectrum except for the bottleneck that renders spectra shallower. The fiducial run R5F is the $1536^3$ DNS. Runs of Figures 1-2 are not included, except for the DNS run R5F and the LES-PH run R5a.}
\begin{ruledtabular}
\begin{tabular}{cccccccccccccc}
Run & Res. & $10^5\nu$ &  $\Omega$  & $\omega_{rms}$  & $T_I$ &$10^2Ro$ &$10^{-3}Re$ & $U_{rms}$ & $L_{0,I}$  & $L_{0,F}$ &  $e$ & $e+h$ & $\Pi_H/[k_F\Pi_E]$ \\
\hline
R1    &  192   & 25.0 &   4.5  & 10.5 & 17 &  4.6   &   8.3 & 0.9 & 2.24 & 3.04 & 2.1 & 4.1 &  1.5  \\ 
R2    &  192   & 25.0 &   9.0  & 11.7 &  8 &  2.9   &   9.3 & 1.1 & 2.10 & 4.30 & 2.2 & 4.2 &  1.7  \\ 
R3    &  192   & 16.0 &   1.8  & 16.0 & 14 & 24.9   &   4.3 & 0.8 & 0.88 & 0.99 & 1.7 & 2.8 &  1.1  \\ 
R4    &  192   & 16.0 &   4.5  & 12.4 & 14 &  6.1   &   9.7 & 0.9 & 1.67 & 3.79 & 1.9 & 3.6 &  1.3  \\ 
\hline
R5F   &  1536  & 16.0 &   9.0  & 15.3 &  9 &  2.9   &  14.7 & 1.1 & 2.14 & 4.32 & 2.0 & 3.9 &  1.9  \\ 
R5    &  192   & 16.0 &   9.0  & 12.8 &  8 &  3.3   &  13.3 & 1.1 & 1.88 & 4.55 & 2.1 & 3.8 &  1.8  \\ 
R5a   &  96    & 16.0 &   9.0  &  8.5 & 10 &  2.7   &  15.3 & 1.1 & 2.23 & 5.21 & 2.2 & 3.8 &  1.9  \\ 
R5z   &  192   & 16.0 &   0.1  & 21.8 &  7 & 1166 &   3.6 & 0.9 & 0.64 & 0.64 & 1.6 & 2.5 &  1.1  \\ 
\hline
R6    &  192   & 16.0 &  18.0  & 10.4 & 10 &  1.6   &  16.7 & 1.2 & 2.16 & 4.58 & 2.1 & 3.9 &  2.5  \\ 
R7    &  192   & 16.0 &  36.0  & 10.8 & 10 &  0.8   &  19.4 & 1.3 & 2.37 & 4.50 & 2.1 & 3.8 &  2.7  \\ 
R8    &  192   & 16.0 & 117.0  & 12.5 &  8 &  0.3   &  14.9 & 1.3 & 1.77 & 2.99 & 2.2 & 4.1 &  2.3  \\ 
R9    &  192   & 11.9 &  18.0  & 11.1 & 10 &  1.4   &  25.0 & 1.2 & 2.41 & 4.76 & 2.1 & 3.9 &  2.3  \\ 
R10   &  192   & 10.2 &  42.4  & 11.7 & 10 &  0.6   &  32.0 & 1.3 & 2.46 & 3.93 & 2.1 & 3.8 &  2.6  \\ 
\hline
R11   &  192   &  8.0 &   9.0  & 14.4 &  8 &  3.8   &  24.9 & 1.2 & 1.72 & 3.79 & 1.9 & 3.3 &  1.7  \\ 
R11a  &  96    &  8.0 &   9.0  &  7.6 & 14 &  1.9   &  44.9 & 1.1 & 3.22 & 5.87 & 2.3 & 4.0 &  1.9  \\ 
\hline
R12   &  192   &  8.0 &  18.0  & 11.1 & 10 &  1.3   &  43.3 & 1.3 & 2.75 & 4.70 & 2.1 & 3.8 &  2.5  \\ 
R13   &  192   &  8.0 &  36.0  & 10.9 & 10 &  0.6   &  47.5 & 1.3 & 2.87 & 4.08 & 2.0 & 3.7 &  2.7  \\ 
R14   &  384   &  8.0 &  36.0  & 15.1 & 10 &  0.9   &  36.6 & 1.4 & 2.14 & 4.18 & 2.1 & 3.9 &  3.2  \\ 
R15   &  384   &  5.3 &  72.0  & 20.9 &  6 &  0.6   &  42.3 & 1.4 & 1.58 & 3.54 & 2.0 & 3.6 &  3.0  \\ 
R16   &  192   &  5.0 &  18.0  & 12.4 &  9 &  1.5   &  62.6 & 1.3 & 2.44 & 4.08 & 2.0 & 3.5 &  2.5  \\ 
R17   &  192   &  4.5 &  36.0  & 15.0 &  6 &  1.0   &  53.8 & 1.4 & 1.79 & 3.91 & 2.0 & 3.4 &  2.8  \\ 
R18   &  192   &  2.5 &  36.0  & 13.9 &  8 &  0.9   & 108.8 & 1.3 & 2.02 & 3.79 & 1.9 & 3.2 &  3.4  \\ 
\hline
R19   &  512   &  1.6 &   9.0  & 22.7 &  8 &  3.7   & 125.5 & 1.2 & 1.74 & 4.16 & 1.8 & 3.4 &  2.1  
\end{tabular} \end{ruledtabular} \label{tab1} \end{table*}

\subsection{The procedure}

Using the LES-PH model tested in \S \ref{s:testing}, 
we now proceed to a parametric study of forced helical rotating turbulence in terms of both the Reynolds and Rossby numbers, examining the scaling laws of energy and helicity in the direct cascade. The Reynolds number varies from 4300 (for smaller values of the Reynolds number it is difficult to ascertain the inertial index of the spectra) to $1.1 \times 10^5$ , and the Rossby number varies from $\approx 12$ to $0.003$. Most runs were done on a grid of $192^3$ points, although both lower and higher resolution runs were performed in order to evaluate to what extent the results depended on the resolution of the LES. 
All runs were started in the same fashion, using the same developed turbulence state with weak rotation ($\Omega =0.06$) as an initial condition, and with the forcing via the ABC flow as given in \S \ref{s:eqs} with $k_F=7$. Data pertinent to the runs can be found in Table \ref{tab1}. In the presence of moderate to strong rotation, an inverse cascade develops in which case the parameters listed in the table were computed at the time of the onset of the inverse cascade $T_I$; note that only two runs (R3 and R5z) do not show an inverse cascade. Note also that $\omega_{rms}$ depends on the resolution since the enstrophy spectrum peaks in the vicinity of the dissipative scale--which is in general not resolved in the LES--for energy spectra shallower than $k^{-2}$ , leading to an under-estimation of the micro Rossby number $\omega_{rms}/2\Omega$ by roughly 50\% in some cases.

We first contrast in Fig. \ref{scatter_tot} the temporal evolution of energy, dissipation and helicity for several runs. Times are indicated in units of eddy turnover time, and as the rotation rate increases, the inertial wave time  decreases in the same manner so that the same physical time on the plots corresponds to a larger number of wave periods. At fixed viscosity (left column), the growth rate of the energy associated with the inverse cascade is rather independent of rotation except for the lowest rotation rate. Otherwise, the time scales are the same for the onset of the cascade itself. This is valid as well at fixed rotation rate and variable $Re$ as seen in the right column, with now comparable energy levels. The discrepancy in the dissipation between the DNS and LES at early times has already been commented upon in the preceding section (see also Fig. \ref{compa_energy}). The dynamics of helicity is very similar for all these runs, with again variations in the amplitude with $Re$ but not so with $Ro$.

The main objective of the parametric study is to measure the spectral indices of energy and helicity, $e$ and $h$, as well as the helicity to energy flux ratio normalized by $k_F$ (see below). To measure $e,h$, a fit is performed in the inertial range of $E(k)$ and $H(k)$ of the form:
\begin{equation}
E(k) \sim k^{-e} , \ \ \ \ \ \ \  H(k) \sim k^{-h} \ .
\end{equation}
The spectra for the fit were computed in three ways: (i) using the spectra at the time of the onset of the inverse cascade $T_I$; (ii) as a time average over an interval $\Delta t=5$ starting from $t=T_I$; and (iii) as an average in the interval $t \in [T_I, 2T_I]$. Although the actual values of the spectral indices were observed to depend slightly on how these estimations were performed, it was observed that a classification based on whether the sum satisfied $e+h \approx 4$ (which will be associated with a wave-dominated regime, and corresponds to the black dots in Fig. \ref{scatter_slope_flux}) or $e+h \not= 4$ (which will be associated with helical turbulence, and corresponds to the gray dots) remains unchanged irrespective of the estimation criteria. The choice finally used to define the spectral indices was thus to perform a temporal average over a not too long time interval (method (ii)) to avoid problems associated with non-stationarity because of the inverse energy cascade and the energy accumulation at large scale. Once the time averaged spectra were obtained, the least squares fit to get the indices was done in the range $k\in[15,45]$. We also performed a spectral fit in the interval $k\in [19,45]$ and $k\in [15,60]$ but saw no measurable difference in the results.

The choice to measure $e+h$ has to do with the two expected regimes for helical rotating turbulence. If rotation is weak, a regime close to Kolmogorov occurs, where both the energy and the helicity cascade directly to small scales with a $k^{-5/3}$ spectrum. In this case, which we refer to as ``helical turbulence'', $e+h=10/3$. When rotation is strong, waves slow-down the energy transfer giving $E(k) \sim k^{-2}$ in the non-helical case (Dubrulle and Valdetarro 1992) and $e+h=4$ in general if helicity is present (see Mininni and Pouquet 2009ab). This case, we refer to as the ``wave-dominated regime''.

An important ingredient for the helical regime $e+h=4$ to occur 
is that the energy flux to small scales be negligible compared to the helicity flux, so that the direct cascade is dominated by helicity. This can be independently verified by computing the energy and helicity fluxes: 
$\Pi_E(K) = \sum_{k=1}^K T_E(k) ,$ and $\Pi_H(K) = \sum_{k=1}^K T_H(k)$.
After averaging over the same time interval used to measure the indices $e$ and $h$, a dimensionless ratio $\Pi_H/(k_F \Pi_E)$ can be computed over the same range of wavenumbers $k\in[15,45]$. When examining this ratio as a function of both the Rossby and Reynolds numbers, it is found that at low rotation rate the ratio is close to unity, as expected, but it is clear that it increases with rotation rate, confirming a previous analysis using DNS (Mininni and Pouquet 2009a); this ratio increases as well as the turbulence strengthens (see Table \ref{tab1}). Thus, one can expect that helical rotating flows in the wave-dominated regime will be characterized by helicity dynamics at small scale.

\begin{figure}[h!] 
\includegraphics[width=8.7cm, height=70mm]{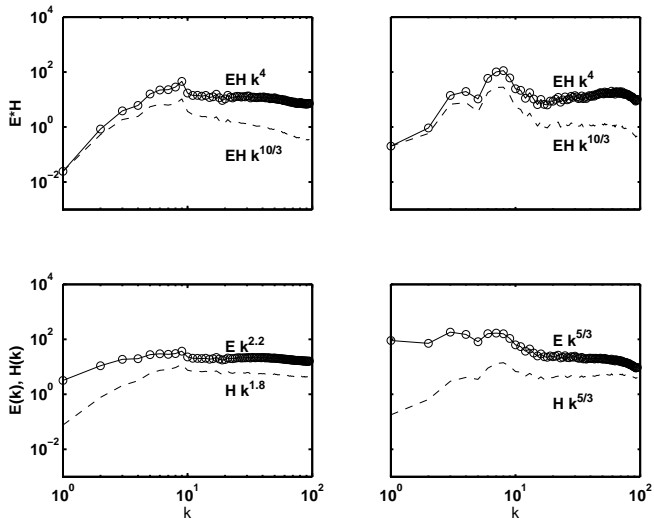}
\caption{{\it Left column:} Compensated spectra for $E(k)H(k)$ for run R1; the bottom plots are the individual spectra $E(k)$ (solid) and $H(k)$ (dash), compensated respectively by $k^{2.2}$ and $k^{1.8}$. {\it Right column:} Same for run R17 (higher rotation and Reynolds number), with below the spectra now both compensated by a Kolmogorov law.}
\label{scatter_spec1}  
\end{figure}

Before proceeding to the details of the parametric study, we give a few examples of the different behaviors that arise as the Reynolds and Rossby numbers are changed. While for most of the flows, the temporal evolution of global statistics is rather similar (see Fig. \ref{scatter_tot}), the resulting spectra in the runs do differ. This is illustrated by Fig. \ref{scatter_spec1}, which gives (top) the product of spectra $E(k)H(k)$ for two different runs compensated by the two laws discussed above: $k^4$ (solid line with circles) and $k^{10/3}$ (dash line). Individual energy and helicity spectra are also shown in the figure for the same two runs (bottom), compensated by laws that are the best fit to the data. Even though the resolution of the LES is modest, the fact that the spectra do not need to display a dissipation range (dissipation being taken care of through the effect of the eddy viscosities) allows for a good determination of spectral indices. All the spectra shown in the figure correspond to a time average from $t=T_I$ to $t=30$. The run on the right of Fig. \ref{scatter_spec1} shows a scaling which seems to be close to the classical Kolmogorov law ($e+h=10/3$), whereas in the run on the left the $e+h=4$ law emerges rather convincingly, with an energy spectrum measurably different from a $e=5/3$ law, the best fit giving $E(k)\sim k^{-2.2}$, $H(k)\sim k^{-1.8}$.

\begin{figure}[h!] 
\includegraphics[width=8.77cm, height=70mm]{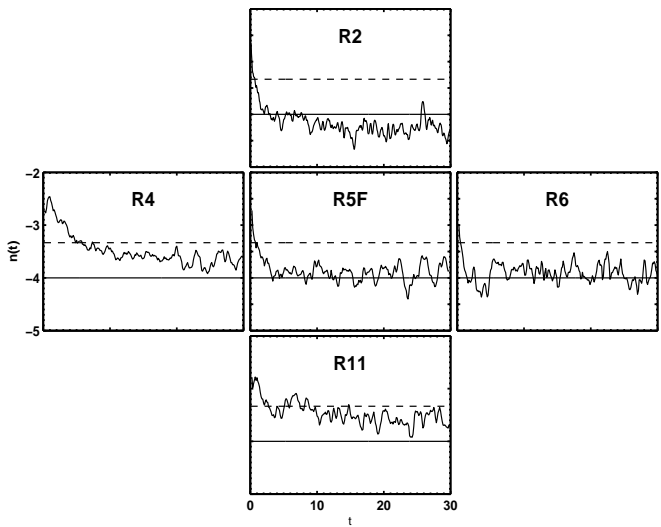}
\caption{Sum of the energy and helicity indices $n(t)=-(e+h)$ as a function of time. R5F is the large DNS and all other cases are LES. Runs 2, 5F and 11 have the same imposed rotation ($\Omega=9$) and decreasing viscosity whereas runs 4, 5F and 6 have the same viscosity ($\nu=1.6 \times 10^{-4}$) and increasing rotation. The resulting dynamics leads to Reynolds and Rossby numbers not quite so well ordered (see Table): for the runs  in the vertical, the Rossby numbers vary by 50\%, and the Reynolds numbers by a factor $\sim 2.3$, whereas for the runs in the horizontal, they vary respectively by a factor 6 and 1.8. Axes have the same scales in all plots; the values of $-4$ and $-10/3$ are shown as solid and dashed horizontal lines.}
\label{scatter_spec2}  
\end{figure}

As previously mentioned, spectral indices vary with time as seen in Fig. \ref{scatter_spec2}, where a fit to obtain $e+h$ was performed in the same range of wavenumbers for the instantaneous spectra as a function of time, using the same five runs as in Figure \ref{scatter_tot}; the solid and dash lines represent respectively the law derived in Mininni and Pouquet (2009a) and the dual Kolmogorov law and are provided for reference. Apart from a slower evolution toward an established spectral law in the case of lower rotation rate (middle row, left-most), the data is compatible with a breaking of universality in helical rotating turbulence; however, other spectral indices cannot be ruled out (observe the middle column). More numerous as well as better resolved runs may be needed to assess this point further; there may be some ambiguity in assessing inertial indices when the runs are close to the transition observed in Fig. \ref{scatter_slope_flux} (see \S \ref{ss:result} below), and furthermore there could be other, as yet unknown, dynamical regimes.

\subsection{The emergence of two regimes} \label{ss:result}

With the previously discussed caveats in mind, we now examine all runs reported in Table \ref{tab1} and classify them based on the resulting sum of spectral indices, $e+h$. A first look at the table indicates that, at fixed Reynolds number, a transition toward the wave-dominated regime takes place as the rotation rate increases, whereas at fixed rotation once the eddy turn-over time becomes smaller than the inertial wave period, Kolmogorov-like turbulence takes over. It is not clear from these simulations whether the solution $e=h=5/3$ is favored, or whether a solution like the one postulated in Brissaud et al. (1973) for helical turbulence emerges, the differences between inertial indices being too minute and the resolution of the simulations not large enough.
 
In order to check consistency of the results, R13 and R14 were run with same $[\nu, \Omega]$ but different resolution; the corresponding Reynolds and Rossby numbers differ respectively by 6\% and 15\%, whereas the flux ratio differs by 18\% (see Table \ref{tab1}). However, these two runs have the same inertial index dynamics. Other convergence tests have been performed that are not reported here, leading to similar conclusions.

A demarcation seems therefore clear: at fixed Rossby number, a higher Reynolds number gives access to smaller scales and shorter turn-over times with the result of having turbulence prevailing over waves and a transition to a state that is consistent with a Kolmogorov-like scaling with dual energy and helicity cascade. Conversely, at fixed Reynolds number, increasing rotation leads to a smaller inertial wave time scale and a prevalence of rotation and of the $e+h\approx 4$ regime found in Mininni and Pouquet (2009a) for helical rotating turbulence. A pure helicity cascade with no energy flux is not observed, the highest value of the ratio of fluxes $\Pi_H/[k_F \Pi_E]$ being $\approx 3.4$. It is also remarkable that, when examining separately the energy and helicity spectra, in several cases we observe that the latter is well defined (either a Kolmogorov law or close to a $k^{-2}$ law), while the index of the energy spectrum may not be so well defined.

\begin{figure}  
\includegraphics[width=8.7cm, height=67mm]{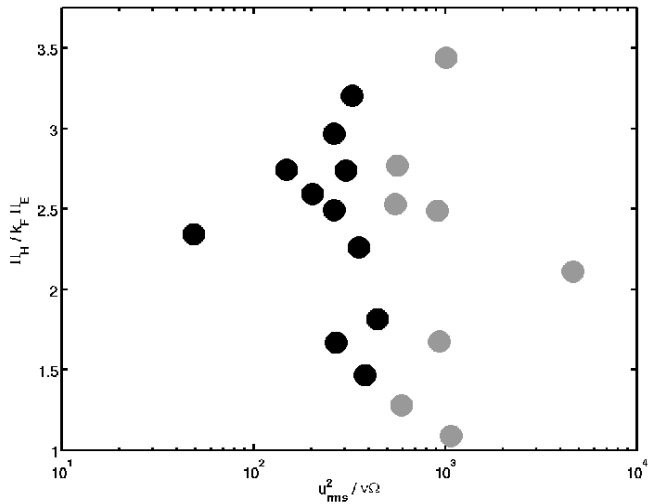} 
 \caption{Scatter plot of the normalized ratio of helicity flux to energy flux as a function of $N_C=Re Ro=U^2_{rms}/[\nu \Omega]$. Note the rather sharp vertical partition around $N_C\sim 500$. 
} \label{scatter_slope_flux}  
\end{figure}

These results are summarized in Fig. \ref{scatter_slope_flux}, which shows the runs that have $e+h \approx 4$ with black dots, and the other runs with grey dots. The two axes correspond to the normalized ratio of the helicity to the energy flux to small scales $\Pi_H/[k_F\Pi_E]$ as a function of the dimensionless parameter $N_C=Re Ro=U^2_{rms}/[\nu \Omega]$. Since the effect of viscosity and rotation are antinomic, their product, properly adimensionalized by the kinetic energy of the system, is a determining parameter. Indeed, the clear demarcation discussed before can be observed as the abrupt partition of the black and grey dots around $N_C \approx 500$. This parameter is easily obtained in the case when the energy spectrum is $E(k)\sim (\epsilon_E \Omega)^{1/2} k^{-2}$ (Dubrulle and Valdettaro 1992, Zhou 1995) by simply searching for the proportionality of the inverse of the dissipation wavenumber that emerges from the above spectrum $k_D\sim [\epsilon_E/(\nu^2 \Omega)]^{1/2}$ and the characteristic large scale of the flow (Canuto and Dubovikov 1997). This is consistent with the idea above that sufficiently small scales must be excited for the eddy turn-over time to be smaller than the inertial wave period and have a transition towards the turbulent regime. Taking helicity   into account, its flux $\epsilon_H$ also comes into play and dimensional analysis becomes undetermined, but the idea remains that the dissipation of energy evaluated using the appropriate energy spectrum where waves play a role should lead to a parameter that governs the dynamics.

This result may seem at odds with the conclusions drawn in Nolan (2005) where the importance of a different combination of dimensionless numbers, namely $Re/Ro=\Omega L^2/\nu_{\ast}$, is emphasized for studying the physics of tornadoes. This latter parameter can be viewed as the ratio of the vortex circulation to the eddy viscosity $\nu_{\ast}$ and is sometimes called the vortex Reynolds number, balancing frictional and advective terms in the boundary layer. However, it should be noted that in our simulations there are no boundary layers (although internal boundary layers develop), and, furthermore, in the presence of strong turbulence, the eddy viscosity that replaces $\nu_{\ast}$ in the above expression becomes proportional to $U^2_{rms}$, and, thus, it may be that in the turbulent case the parameter $Re Ro=U^2_{rms}/(\nu \Omega)$ becomes the relevant parameter. Moreover, note that the purpose of our parametric study is quite different: whereas Nolan (2005) seeks a predictive parameter before a tornado forms, we are here dealing with the overall scaling properties of small scale fluctuations in a rotating turbulent flow with updrafts or downdrafts (provided by the helicity).

Besides different scaling laws, the other measurable differences we found between the two regimes identified in this paper, at fixed rotation rate, are a longer turn-over time $\tau_{NL}$ in the case when $e+h=4$ together with a slightly slower growth rate in the inverse cascade, due presumably to a more efficient Beltramization of small scales because of the excess of small scale helicity. It would be of interest to investigate in detail the structure of strong laminar columns that form in these flows (Beltrami Core Vortices, see Mininni and Pouquet 2009c), and relate them to their far-field environment, but this is left for future work. However, a preliminary examination of these structures does not reveal any marked difference of behavior between the two regimes: the same spatial juxtaposition of persistent laminar large columnar vortices elongated along the axis of rotation, and a web of intense small-scale vortices is visible in both cases (Mininni and Pouquet 2009bc). These structures are different from what happens in a non-helical flow (Mininni et al. 2008) in which case no laminar columnar vortices appear in the turbulent regime; this may indicate that, whatever the spectral law, helicity plays an important role in determining the statistical properties of rotating helical turbulence and the stability of its structures.

Another remark that can be made at this point is that, as in all that precedes, the emphasis was on the importance of helicity, which is a topological invariant measuring the knottedness of vortex lines (Moffatt 1969), but no computations were considered of rotating turbulence without net helicity. A continuation of this work will obviously involve performing computations with forcing possessing different degrees of relative helicity. In fact, in the non-helical case of a Taylor-Green forcing studied in Mininni et al. (2008), a $k^{-2}$ spectrum was found, and one may ask whether it will also disappear in favor of a pure Kolmogorov spectrum as the Reynolds number is increased for fixed Rossby number. This point is left for future investigation.

\section{Discussion and conclusion} \label{s:conclu}

A significant effort has been put into modeling turbulent flows, both in the engineering and atmospheric contexts, as well as in astrophysics. What the results presented in this paper show is the fact that the helical model developed previously in Baerenzung et al. (2008a) works reasonably well when compared to high resolution high Reynolds number and moderate Rossby number direct numerical simulations of helical rotating turbulence. In this particular case, the inclusion of helicity improves the results, whereas in the non rotating case with both non-helical or helical forcing, or in the rotating case with non-helical forcing, the inclusion of helicity in the model neither enhances nor degrades the results. For the same problem, the Chollet-Lesieur (1981) model does not obtain the correct growth rate of energy in the inverse cascade; it can be viewed as somewhat deficient insofar as it seems too dissipative. This can be linked to the fact that, in the presence of strong helicity, the nonlinearities are damped (Kraichnan and Panda 1988) and thus the turbulent dissipation is substantially diminished when compared to the non-helical case, particularly so in the rotating case (Teitelbaum and Mininni 2009).

The CPU and memory usage savings when computing with the model are impressive, since a grid of $96^3$ points was used in all tests of the LES against the DNS run on a grid of $1536^3$ points for the same Reynolds and Rossby numbers. It is possible that further gain may be obtained when comparing to DNS performed on larger grids at higher Reynolds numbers, but such an evaluation of the optimum gain in a LES will have to wait for petascale computers and beyond. Indeed, a minimum of inertial range has to be resolved in the LES computation in order to compute the eddy diffusivity and eddy noise expressed in terms of the resolved energy and helicity spectra. Another advantage of the LES model is that doing temporal three-dimensional visualization of flows at resolution of $1024^3$ grid points and above is still a costly exercise, demanding in memory usage as well as computer and human time. The LES runs, from that point of view, give a proxy visualization of the flow at almost no cost and are potentially very useful even when errors in some of the global statistical properties of the flow may be as large as 10\%, as observed in some of the LES test runs on grids of $96^3$ points.

A comparison with an under-resolved DNS on a grid of $160^3$ points showed that all LES perform significantly better. Both the energy evolution and its spectral distribution are wrong for the under-resolved DNS: there is insufficient dissipation and the inverse energy cascade is too vigorous with an accumulation of energy at the largest scale as well as in the smallest scales, while the spectrum at intermediate scales is greatly under-valued. There is no doubt that at an equivalent resolution and/or numerical cost, the several LES tested here work significantly better. Finally, we note that the LES-PH model has twice the cost in computational time compared to a DNS at the same grid resolution. 

An obvious application of the helical model tested here for rotating flows is to explore regimes unattainable today with DNS. We have done so in the second part of this paper, examining the scaling laws of energy and helicity in rotating flows in the direct cascade range. We find that there can be two regimes, one where inertial waves interact with helical eddies in the limit of low Rossby number, yielding energy spectra steeper than the Kolmogorov law, and one close to the Kolmogorov regime.

Several questions need to be examined in the future, among which are: What are the structures in the regime that obtains at large Reynolds number? Are the properties of fully developed turbulent flows at low Rossby number close to the predictions of weak turbulence theory (Galtier 2003)? How does  the inverse cascade scale, and how does it saturate, when sufficient resolution is present at large scale? What will be the effect of adding a friction term at large scale?
 What is the effect of a moderate amount  of helicity 
 (only the cases of maximal or zero helicity forcing have been tested up to now)? Would a different choice of forcing (such as a two-dimensional force, or a random force) affect the results? 

Furthermore, the model presented in this paper does not include anisotropies or inhomogeneities of the small scales, nor does it take into account bottom topography (see e.g., Fredriksen 1999) or other realistic physics relevant when dealing with atmospheric flows. Memory effects are neglected as well in our modeling since the EDQNM on which it is based is a Markovianized closure. In that light, a stochastic approach as developed by several authors (see e.g. Majda et al. 2003, Delsole 2004, Fredriksen and Keppert 2006) has led to significant progress in the modeling of turbulent flows as tested against direct numerical simulations at moderate resolutions.
Further improvements and tests will be needed to capture as well the memory effects of turbulent flows.
 Another obvious drawback of closure models of turbulence such as the EDQNM is that all information on moments of the stochastic velocity field above second-order is lost, and phase information among Fourier modes is lost as well. Thus,
for example, intermittency is not present in the EDQNM, although it is observed in the EDQNM-based LES since the LES, in principle, captures sufficient information on the structure of the inertial range.
Such improvements will require non-trivial developments. On the other hand, adding to the dynamics one or several scalar fields, such as the potential temperature and the water vapor, cloud water and rain water mixing ratios as done in Wicker and Wilhelmson (1995), passively advected by the flow, is not necessarily that cumbersome, since the EDQNM for the passive scalar problem has been written 
and thus the transport coefficients are known. 
Using adaptive mesh refinement in the presence of boundaries, as done in Wicker and Wilhelmson (1995), or possibly with spectral accuracy (Rosenberg et al. 2006, 2007) will also enhance our capacity to analyze complex flows.

Some physical models have also incorporated helicity on the dynamics (Lautenschlager et al. 1988, Yokoi and Yoshizawa 1993, Li et al. 2006). For example, motivated by observations of tropical cyclones giving estimates for the averaged helicity in a variety of flows (Anthes 1982, Etling 1985, Lilly 1986), a helical subgrid-scale parametrization was proposed in Lautenschlager et al. (1988), following similar studies in magnetohydrodynamics (see Krause and R\"udiger 1974 for the case of neutral fluids). These analyses differ from the present study insofar as they concentrate on large-scale instabilities, whereas this paper has been devoted to the issue of small-scale statistics. For example, in Lautenschlager et al. (1988), the modeling, backed up by low-resolution DNS, writes in terms of transport coefficients proportional to the vorticity, 
inspired from the equivalent destabilizing effect of small-scale helicity on large-scale magnetic fields in MHD. 
Instabilities involving three derivatives of the velocity (in Fourier space, $\sim k^3 \hat v({\bf k},t)$) in the presence of helicity were computed in Pouquet et al. (1978) using the renormalization group when considering the limit $k\rightarrow 0$, in which case they are sub-dominant. The problem remains of properly modeling the effect of small-scale helicity on large-scale flows, the approach taken in this paper being a modification to the eddy viscosity integrating a helical component, although large scale instabilities may develop as well (Frisch et al. 1984).

Finally, our model is isotropic in the unresolved scales, an assumption that can of course be relaxed (see Cambon and Scott 1999, Sagaut and Cambon 2008 and references therein). It would lead to more complex expressions for the energy transfer terms involving variations of correlation functions in terms of both $k_{\perp}$ and $k_{\parallel}$ (where $\perp$ and $\parallel$ refer to the direction of rotation).
This implies that a numerical integration of such anisotropic closures is significantly more costly that what is performed in the work presented here, since angles as well as wavenumbers have to be discretized. As shown above, the isotropy assumption works relatively well at the moderate Rossby numbers tested here, which are close to atmospheric values although at substantially lower values of the Rossby number, more complex models may have to be developed.

\begin{acknowledgments}
The large computation on a grid of $1536^3$ points was done through a special allocation within the ADS program at NCAR which is sponsored by NSF. PDM is a member of the Carrera del Investigador Cient\'{\i}fico of CONICET.
\end{acknowledgments}



\begin{thebibliography}{99}

 
 \bibitem{andre_lesieur}
Andr\'e, J.C. and Lesieur, M., 1977:
Influence of Helicity on the  Evolution of Isotropic Turbulence at High Reynolds Number.
{\it J. Fluid Mech.}, {\bf 81}, 187-207.

\bibitem{anthes82}
Anthes, R., 1982: Tropical cyclones, their evolution, structure and effects. {\it Meteorological Monographs} {\bf 19}, Number 41. American Meteorological Society.

 \bibitem{LESPH}
Baerenzung, J., Politano, H., Ponty, Y., and Pouquet, A., 2008a:
``Spectral Modeling of Turbulent Flows and the Role of Helicity,''
  {\it Phys. Rev. E}, {\bf 77}, 046303.
  
   \bibitem{model_MHD}
------, 2008b:
``Spectral Modeling of Magnetohydrodynamic Turbulent Flows,''
{\it Phys. Rev. E} {\bf 78}, 026310.

 \bibitem{LES_ROT}
------,  Mininni, P.D., Politano, H.,  Ponty, Y., and Pouquet, A., 2009:
``Spectral Modeling of Rotating Turbulent Flows,''
submitted to  {\it Phys. Rev. E}. See also arXiv:0812.1821v1.

\bibitem{brissaud}
Brissaud, A., Frisch, U., L\'eorat, J.,  Lesieur, M., and Mazure, A., 1973: Helicity Cascades in Fully Developed Isotropic  Turbulence.
{\it Phys. Fluids}, {\bf 14}, 1366-1367.

\bibitem{cambon_scott} 
Cambon, C., and J.F. Scott, J.F., 1999:
Linear and Non-Linear Models of Anisotropic Turbulence.
{\it Ann. Rev. Fluid Mech.}, {\bf 11}, 1--53.

\bibitem{canuto}
Canuto, V.M., and Dubovikov,  M.S., 1997:
A dynamical model for turbulence. V. The effect of rotation.
{\it Phys. Fluids}, {\bf 9}, 2132-2140.

\bibitem{chen05}
Chen, Q., Chen, S., Eyink, G. and Holm, D., 2005: Resonant interactions in rotating homogeneous 
three-dimensional turbulence. {\it J. Fluid Mech.} {\bf 542}, 139-164.

\bibitem{chollet}
Chollet, J.P.,  and Lesieur, M., 1981: 
Parametrization of small scales of three-dimensional isotropic turbulence utilizing spectral closures.   
 {\it J. Atmos. Sci.}, {\bf 38}, 2747-2757.
 
 \bibitem{delsole}
DelSole, T., 2004: Stochastic Models of Quasigeostrophic Turbulence. {\it Surv. Geophys.}, {\bf25}, 107Ð149.

\bibitem{dubrulle}
Dubrulle, B. and Valdetarro, L., 1992:
Consequences of rotation in energetics of accretion disks.
{\it Astronom. Astrophys.} {\bf 263}, 387-400.

\bibitem{etling85}
Etling, D., 1985: Some aspects of atmospheric flows. {\it Contrib. Atmosph. Phys.} {\bf 58}, 88-100.

\bibitem{australian2}  Frederiksen, J. S., 1999: Subgrid-Scale Parameterizations of Eddy-Topographic Force, Eddy Viscosity, and Stochastic Backscatter for Flow over Topography. {\it J. Atmos. Sci.}, {\bf 56}, 1481-1494.

\bibitem{australian1}  ----, and S.M. Kepert, 2006: Dynamical Subgrid-Scale Parameterizations from Direct Numerical Simulations. {\it J. Atmos. Sci.}, {\bf 63}, 3006-3019.

\bibitem{AKA}
Frisch, U., Scholl, H., She, Z-S. and Sulem, P-L., 1984: A new large-scale instability in three-dimensional incompressible flows lacking parity invariance, {\it Fluid Dyn. Res.} {\bf 3}, 295--298.

 \bibitem{galtier03} Galtier, S. 2003: Weak Inertial-Wave Turbulence Theory. {\it Phys. Rev. E} {\bf 68}, 015301(R).

\bibitem{holm_kerr}
Holm, D.D.,  and Kerr, R., 2002: Transient Vortex Events in the Initial Value Problem for Turbulence.
{\it Phys. Rev. Lett.} {\bf88}, 244501.

\bibitem{hunt}
Hunt, J.C.R., and F. Hussain, 1991:
A note on velocity, vorticity and helicity of inviscid 
fluid elements. {J. Fluid Mech.} {\bf 229}, 569-587.

\bibitem{farge_01}
Jacobitz, F. G., Liechtenstein, L., Schneider, K. and  Farge, M., 2008:
On the structure and dynamics of sheared and rotating turbulence: direct numerical simulation and wavelet-based coherent vortex extraction. {\it Phys. Fluids} {\bf 20}, 045103.

\bibitem{kane_08}
O'Kane, T.J., and Frederiksen, J. S., 2008:
A comparison of statistical dynamical and ensemble prediction methods during blocking.
{\it J. Atmos. Sci.}, {\bf 65}, 426-447.

\bibitem{kraichnan73}
Kraichnan, R.H., 1973: Helical turbulence and absolute equilibrium.
{\it J. Fluid Mech.} {\bf 59}, 745-752.

\bibitem{kraichnan_1976}
-------, 1976:
Eddy viscosity in two and three dimensions.
{\it J. Atmos. Sci.}, {\bf 33}, 1521-1536.

\bibitem{kraichnan_panda}
--------,  and Panda, R., 1988:
Depression of Nonlinearity in Decaying Isotropic Turbulence.
{\it Phys. Fluids}, {\bf 31}, 2395-2397.

\bibitem{krause74}
 Krause, F.  and  R\"udiger, G., 1974:
 On the Reynolds stress in mean-field hydrodynamics I. Incompressible homogeneous isotropic turbulence.
 {\it Astron. Nachr.} {\bf  295}, 93.

\bibitem{krstu}
Krstulovic, G.,  Mininni, P., Brachet, M., and Pouquet, A., 2009:
Cascades, Thermalization and Eddy Viscosity in Helical Galerkin Truncated Euler Flows.
submitted.

\bibitem{kurien}
Kurien, S., Taylor, M.A., and Matsumoto, T.,  2004:
Cascade Time Scales for Energy and Helicity in Homogeneous Isotropic Turbulence .
{\it Phys. Rev. E}, {\bf 69}, 066313.

\bibitem{lauten88}
Lautenschlager, M., Eppel, D.P. and Thacker, W.C., 1988:
Subgrid parametrization in helical flows.
{\it Beitr. Phys. Atmosph.} {\bf 61}, 87-97.

 \bibitem{leith}
 Leith, C.E., 1971: Atmospheric Predictability and
Two-Dimensional Turbulence. 
{\it J. Atmos. Sci.}, {\bf 28}, 145-161. 


\bibitem{li06}
Li,Y.,  Meneveau, C.,  Chen, S. and  Eyink, G.L., 2006:
Subgrid-scale modeling of helicity and energy dissipation in helical turbulence.
{\it Phys. Rev. E} {\bf 74}, 026310.

\bibitem{lilly86}
Lilly, D., 1986: The Structure, Energetics and Propagation of Rotating Convective Storms. Part II: Helicity and Storm Stabilization.
{\it J. Atmos. Sci.} {\bf 43}, 126-140.

\bibitem{majda}
Majda, A. J., Timofeyev, I., and Vanden-Einden, E.  2003: Systematic Strategies for Stochastic Mode Reduction in Climate. 
{\it J. Atmos. Sci.}, {\bf 60}, 1705Ð1722.

\bibitem{markowski}
Markowski, P.M., Straka, J.M.,  Rasmussen, E.N., and Blanchard, D.O., 1998:
Variability of Storm-Relative Helicity during VORTEX .
{\it Monthly weather Rev.}, {\bf 126}, 2959-2971.

\bibitem{markowski_2003}
----,  Straka, J.M., and Rasmussen, E.N., 2003:
Tornadogenesis Resulting from the Transport of Circulation by a Downdraft: Idealized Numerical Simulations.
{\it J. Atmos. Sci.}, {\bf 60}, 795-821.

\bibitem{meneveau_katz}
Meneveau, C., and Katz, J., 2000:  Scale-Invariance and Turbulence Models for Large-Eddy Simulation. {\it Ann. Rev. Fluid Mech.}, {\bf 32}, 1-32.

\bibitem{mininni_rot_TG}
Mininni, P.D., Alexakis, A.,  and Pouquet, A., 2008:
Scale Interactions and Scaling Laws in Rotating Flows at Moderate Rossby Numbers and Large Reynolds Numbers,
submitted to {\it Phys. Fluids} , see also http://arxiv.org/abs/0802.3714 .

\bibitem{mininni_rot_hel}
------, and Pouquet, A., 2009a:
 Helicity Cascades in Rotating Turbulence,''
{\it Phys. Rev. E} {\bf 79}, 026304.

\bibitem{mininni_rot_hel2}
------, and Pouquet, A., 2009b:
 ``Rotating helical turbulence. Part I. Global evolution and spectral behavior''.
    Submitted to {\it Phys. Rev. E},   see also arXiv:0909.1272.
    
\bibitem{mininni_rot_hel3}
-------,  and Pouquet, A., 2009c:
``Helical rotating turbulence. Part II. Intermittency, scale invariance and structures''.
    Submitted to {\it Phys. Rev. E},   see also arXiv:0909.1275.
    
\bibitem{moffatt69}
Moffatt, H.K., 1969:
The degree of knottedness of tangled vortex lines. {\it J. Fluid Mech.} {\bf 35}, 117--129.

\bibitem{moffatt_tsinober}
Moffatt, H.K., and Tsinober A., 1992:
Helicity in Laminar Turbulent flow. {\it Ann. Rev. Fluid Mech.}, {\bf 24}, 281-312.
 
\bibitem{nolan_JAS_2005}
Nolan, D.S., 2005:
A new scaling for Tornado-Like Vortices.
{\it J. Atmos. Phys.}, Notes and Correspondence, {\bf 62}, 2639-2645.

\bibitem{nolan_JAS_1999}
-------,  and Farrell, B.F., 1999:
The Structure and Dynamics of Tornado-Like Vortices.
{\it J. Atmos. Phys.}, {\bf 56}, 2908-2936.

\bibitem{orszag}
Orszag, S.A., 1977: Les Houches Summer School, J.L. Peube and R. Baliand Eds., Gordon and Breach.

\bibitem{palmer}
Palmer, T. N., 2001: A Nonlinear Dynamical Perspective on Model Error: A Proposal for Non-Local Stochastic-Dynamic 
Parameterization in Weather and Climate Prediction Models., {\it  Quart. J. Roy. Meteor. Soc.}, {\bf 127}, 279Ð304. 

\bibitem{pouquet_RNG}
Pouquet, A.,  Fournier, J.D. and Sulem, P. L., 1978: Is helicity relevant
for large scale steady state three--dimensional turbulence?  {\it J. Phys.
Lettres} (Paris), {\bf 39}, L 199--203.

\bibitem{duane_JCP}
Rosenberg, D., Fournier, A.,  Fischer, P., and Pouquet, A., 2006:
Geophysical-Astrophysical Spectral Element Adaptive Refinement
(GASpAR): Object-Oriented h-Adaptive Code for Geophysical Fluid Dynamics
Simulations. {\it J. Comp. Phys.} {\bf 215} 59--80.

\bibitem{duane_MHD}
----, Pouquet , A., and  Mininni, P.D.,  2007:
Adaptive Mesh Refinement with Spectral Accuracy for Magnetohydrodynamics in Two Space Dimensions.
 {\it New J. Phys.}, {\bf 9}, 304. 

\bibitem{rotunno_84}
Rotunno, R., 1984:
An Investigation of a Three-Dimensional Asymmetric Vortex.
{\it J. Atmos. Sci.}, {\bf 41}, 283- 298.

\bibitem{rotunno_85}
--------, R., and Klemp, J., 1985:
On the rotation and propagation of simulated supercell thunderstorms.
{\it J. Atmos. Sci.}, {\bf 42}, 271- 292.

\bibitem{cambon_book}
Sagaut, P., and C. Cambon, 2008:
\newblock {\it Homogeneous Turbulence Dynamics}.
\newblock Cambridge Univ. Press. xxx pp.

\bibitem{smago}
Smagorinsky, J.S., 1963:  General Circulation Experiments with the Primitive 
Equations: I. The Basic Experiment, {\it Mon. Weather Rev.}, {\bf 91}, 99-164.

\bibitem{smith96}
Smith, L., Chasnov, J. and Waleffe, F., 1999:
Crossover from Two- to Three-Dimensional Turbulence.
{\it Phys. Rev. Lett.}, {\bf 77}, 2467-2470.

\bibitem{smith99}
-----, and Waleffe, F., 1999:
Transfer of energy to two-dimensional large scales in forced, rotating three-dimensional turbulence.
{\it Phys. Fluids}, {\bf 11}, 1608-1622.

 \bibitem{smith05} 
 -----, and Lee, Y., 2005: On near resonances and symmetry breaking in forced rotating flows at moderate Rossby number. {\it J. Fluid Mech.} {\bf 535}, 111-142.

\bibitem{pablo_student}
Teitelbaum, T. and Mininni, P.D., 2009:
Effect of helicity and rotation on the free decay of turbulent flows. To appear, {\it Phys. Rev. Lett.}.

\bibitem{tran}
Tran, C. and Bowman, J. 2003 
On the dual cascade in two-dimensional turbulence.
{\it Physica D} {\bf 176}, 242--255.

\bibitem{wicker_wilhelmson_1995}
Wicker, L.J., and Wilhelmson, R.B., 1995:
Simulation and Analysis of Tornado Development and Decay Within a Three-Dimensional Supercell Thunderstorm.
{\it J. Atmos. Sci.}, {\bf 52}, 2675- 2703.

\bibitem{yokoi93}
Yokoi, N. and Yoshizawa, A., 1993:
Statistical analysis of the effects of helicity in inhomogeneous turbulence.
{\it Phys. Fluids}, {\bf A5}, 464--477.

\bibitem{Zhou95}
Zhou, Y., 1995: A Phenomenological Treatment of Rotating Turbulence.
{\it Phys.\ Fluids} {\bf 7}, 2092-2094.

\end{thebibliography}
\end{document}